\documentclass[aps,prd,twocolumn,superscriptaddress,showpacs]{revtex4-1}

\usepackage{graphicx}

\begin{document}


\title{Congeniality Bounds on Quark Masses from Nucleosynthesis}


\author{M.~Hossain~Ali}
\email[]{m\_hossain\_ali\_bd@yahoo.com}
\affiliation{Department of Applied Mathematics, Rajshahi University, Rajshahi 6205, Bangladesh}
\author{M.~Jakir~Hossain}
\email[]{mjakir\_bd@yahoo.com}
\affiliation{Department of Mathematics, Rajshahi University, Rajshahi 6205, Bangladesh}
\author{Abdullah Shams Bin Tariq}
\email[Corresponding author: ]{asbtariq@ru.ac.bd}
\affiliation{Department of Physics, Rajshahi University, Rajshahi 6205, Bangladesh}


\date{\today}

\begin{abstract}
The work of Jaffe, Jenkins and Kimchi [Phys.~Rev.~{\bf D}79, 065014 (2009)] is revisited to see if indeed the region of congeniality found in their analysis survives further restrictions from nucleosynthesis. It is observed that much of their congenial region disappears when imposing conditions required to produce the correct and required abundances of the primordial elements as well as ensure that stars can continue to burn hydrogen nuclei to form helium as the first step in forming heavier elements in stellar nucleosynthesis.  The remaining region is a very narrow slit reduced in width from around 29 MeV found by Jaffe {\it et al.}~to only about 2.2 MeV in the difference of the nucleon/quark masses.  Further bounds on $\delta m_q /m_q$ seem to reduce even this narrow slit to the physical point itself.
\end{abstract}


\maketitle



A reasonably contemporary approach is to study, even without going into anthropic arguments, the nature of alternative universes as one changes the values of physical parameters.  In the parameter space, one then looks for regions that could be similar to our universe and may possibly be congenial to the creation and sustenance of intelligent life \cite{Wei89,Teg97,Teg98a,Teg98b,Hog00,Teg06,Jaf09,Don10}.  Bounds thus obtained may be referred to as congeniality bounds.

In a recent work, Jaffe, Jenkins and Kimchi \cite{Jaf09} studied how sensitive our universe would be to variations of quark masses. For this they chose to study the variations of masses of the three lightest quarks $u$, $d$ and $s$, under the constraint that the sum of these masses, $m_T$ remained fixed. They also studied variations of $m_T$. 

Their basic idea was to find the two lightest baryons for any quark mass combination and consider them to play the roles of the proton and neutron in forming nuclei. In this process they also considered $\Lambda_{\mathrm{QCD}}$ to be an adjusted free parameter that they tuned to keep the average nucleon mass at 940 MeV. They then studied the variation of nuclear stability and, in light of this, tried to obtain the regions of the parameter space where nuclear chemistry in a somewhat familiar form could be sustained.  

\begin{figure}[b]
\includegraphics[width=0.8\columnwidth]{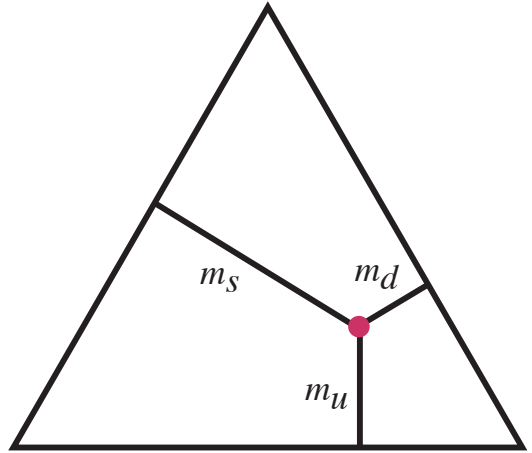}%
\caption{\label{mtriangle}The model space of light quark masses for a fixed $m_T$ shown in the form of a triangle where the distance from the three sides give the three masses.  Figure reproduced from \cite{Jaf09}.}
\end{figure}

The starting point is that the three light quark masses would be changed keeping their sum $m_T$ fixed. This parameter space can be neatly shown in the form of an equilateral triangle [Fig.~\ref{mtriangle}] where the distances of a point from the base and right and left sides are, respectively, the masses of the up, down and strange quarks.

\begin{figure}
\includegraphics[width=0.8\columnwidth]{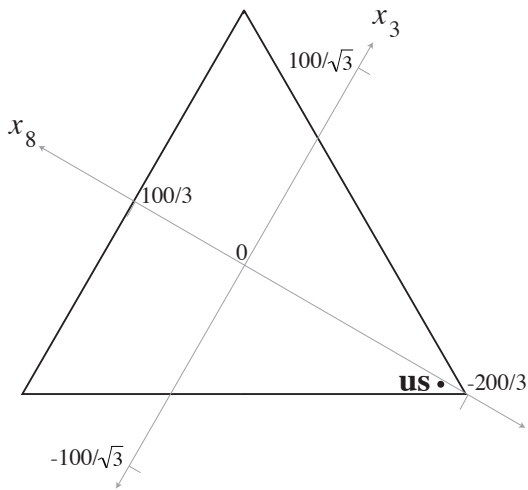}%
\caption{\label{x3x8}The model space of light quark masses parametrized in terms of $x_3$ and $x_8$ reproduced from \cite{Jaf09}.  The point labeled `us' points to the physical value in our present universe and therefore has coordinates $(x_3^\oplus, x_8^\oplus)$.}
\end{figure}

In this manner they identified congenial regions in a triangle parametrized in terms of $x_3$ and $x_8$ [Fig.~\ref{x3x8}] defined as
\begin{eqnarray}
x_3 &=& \frac{2m_3}{\sqrt{3}}\frac{100}{m_T^\oplus}=\frac{100\left(m_u-m_d\right)}{\sqrt{3}m_T^\oplus}\\
x_8 &=& \frac{2m_8}{\sqrt{3}}\frac{100}{m_T^\oplus}=\frac{100\left(m_u+m_d-2m_s\right)}{\sqrt{3}m_T^\oplus}
\end{eqnarray}

\begin{figure}[t]
\includegraphics[width=0.8\columnwidth]{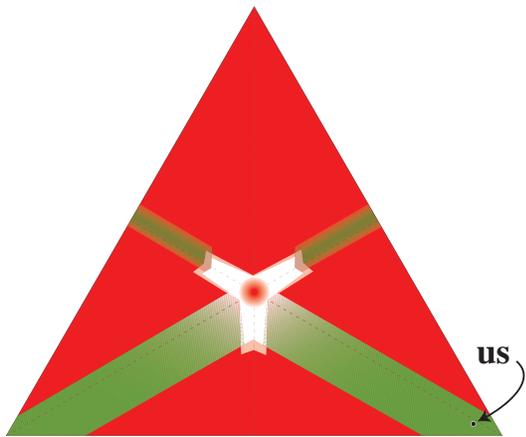}%
\caption{\label{cong}(color online) Figure reproduced from \cite{Jaf09} identifying congenial regions in the quark mass triangle with green bands.  The red and white regions are uncongenial and uncertain, respectively.}
\end{figure}

Here $m_T^\oplus$ is the sum of the light ($u$, $d$ and $s$) quark masses in our universe.  It is obvious that $x_3$ basically gives the isospin splitting while $x_8$ is related to the breaking of $SU(3)_f$ due to the mass of the strange quark. Their results can be summarized in Fig.~\ref{cong}, where the congenial regions are indicated in green \footnote{If you are reading a black and white print, then green appears as lightly shaded and red as deep shaded}. This triangle is for $m_T^{\oplus}$, {\it i.e.}~with $m_T$ as it is in the present universe. They have also studied variations in $m_T$, but our work is limited to commenting on the case of $m_T^\oplus$, understanding that the same arguments qualitatively extend to other values of $m_T$.  This is further justified in the discussions near the end of this report.

It is increasingly being understood that if there is complexity, fine-tuning is inevitable \cite{Brad11}.  Even if one is not happy with anthropic arguments, we simply cannot get away from fine-tuning.  With this in mind, the first impression that one has from Fig.~\ref{cong}, is that the congenial region seems to be surprisingly large - allowing, around one order of magnitude variations in the quark masses. Though this already involves intricate compensating adjustments in $\Lambda_{\mathrm{QCD}}$ to keep the average `nucleon' mass fixed. 

However, one should appreciate the difficulty in setting up a new framework in which a problem can be studied. From this perspective the authors of \cite{Jaf09} should be commended for presenting, literally from scratch, a setup for studying the congeniality bounds on quark masses. This setup can be extended removing some of the constraints used in any further work.  Indeed, considering the significance of the work it was chosen first for a Viewpoint article in Physics \cite{Per09} and then went into a cover story in Scientific American \cite{Jen10}.

In our work, we remain within the provided setup, but extend the analysis to bounds provided by nucleosynthesis. It should be noted here that whereas, on the one hand, nuclear masses and stability expectedly vary comparatively slowly with quark masses; on the other hand, the observed abundances of the lightest nuclei hydrogen and helium provide much more stringent bounds on the variation of nucleon masses. We report below how the congeniality triangle of Fig.~\ref{cong} is modified by the application of these constraints.

At the outset, the variation of the octet baryon masses as one traverses along the borders of the triangle (Figs.~10 and 11 of \cite{Jaf09}) were reproduced to gain confidence in our code and our understanding of the framework.  The fitted parameter values of $c_T$, $c_3$ and $c_8$ from Table III of Ref.~\cite{Jaf09} were used in the equation
\begin{equation}
M_B = C_0 + c_Tx_T + c_8x_8 + c_3x_3 + \left\langle B \left|H_{\mathrm{EM}}\right| B \right\rangle .
\end{equation}
This leads to  
\begin{eqnarray}
M_p = C_0 + 3.68\: x_T + 3.53\: x_8 + 1.24\: x_3 + 0.63\\
M_n = C_0 + 3.68\: x_T + 3.53\: x_8 - 1.24\: x_3 - 0.13.
\end{eqnarray}
The quantity occurring most in the analysis below being
\begin{equation}
M_n-M_p = -2.48\: x_3-0.76.
\end{equation}

It may be noted here that using updated values of baryon masses from the Particle Data Book changes the parameters very slightly and this is neglected considering the qualitative nature of this work.  After a clarification on the adjustment of $C_0$ (corresponding to an adjustment of $\Lambda_{\mathrm{QCD}}$) from the authors \cite{JJK} it was possible to reproduce the figures.

Then the issue of further bounds from nucleosynthesis were studied.  It is well known that the observed abundances of the primordial nuclei hydrogen (protons), helium (alpha particles) etc.~are sensitively tied to the masses of the nucleons \cite{barrowtipler, Hog00}.  The slight difference in the masses of the proton and neutron are responsible for the survival of protons with the observed abundance.  The (un)congenial regions of the triangle is explored further under these constraints.

There are three cases that arise here:

\subsection*{Case I: $x_3 > x_3^\oplus$}

Let us concentrate on the region on the upper right of the triangle with $x_3$ values greater than at the point labeled `us' on the right hand side of the triangle. 

Of the two nucleons, the neutron is heavier by about 1.3 MeV.  If it was just 0.8 MeV less, that would bring it below the electron capture threshold for protons, i.e. it would become energetically favourable for protons to capture electrons and become neutrons.  All the protons would have been converted to neutrons in the Big Bang. The Universe would be full of neutrons and nothing else. We would not be here.  In words of Barrow and Tipler \cite{barrowtipler}
\begin{quote}
Without electrostatic forces to support them, solid bodies would collapse rapidly into neutron stars or black holes. Thus, the coincidence that allows protons to partake in nuclear reactions in the early universe also prevents them decaying by weak interactions. It also, of course, prevents the 75\% of the Universe which emerges from nucelosynthesis in the form of protons from simply decaying away into neutrons. If that were to happen no atoms would ever have formed and we would not be here to know it. [Ref.~\cite{barrowtipler}, p.~400]
\end{quote}

The same issue is also discussed by Hogan \cite{Hog00}
\begin{quote}
The $u$-$d$ mass difference in particular attracts attention because the $d$ is just heavier enough than $u$ to overcome the electromagnetic energy difference to make the proton ($uud$) lighter than the neutron ($udd$) and therefore stable. On the other hand, if it were a little heavier still, the deuteron would be unstable and it would be difficult to assemble any nuclei heavier than hydrogen.
\end{quote}

Therefore, it is necessary to have
\begin{equation}
M_n-M_p \geq 0.5\:\mathrm{MeV}.
\end{equation}
This reduces the congenial region on the upper right of the physical point to 
\begin{equation}
x_3 \leq -0.51.
\end{equation}

\subsection*{Case II: $x_3 < x_3^\oplus$}

Now let us move to the bottem-left side of the triangle (left of the $x_8$ axis) where $x_3$ values are smaller than that at the `us'-labeled point, again concentrating on the right hand side of the triangle.

The key reaction by which hydrogen ‘burns’ in stars such as the sun involves the reaction
\begin{eqnarray}
p+p&\rightarrow& d+e^++\nu+0.42\,\mathrm{MeV}\\
e^++e^-&\rightarrow& 1\:\mathrm{MeV}
\end{eqnarray}
So the total amount of energy released in this reaction is 1.42 MeV.

If the neutron mass was 1.42 MeV (0.15\%) more than it is, this reaction would not happen at all. It would need energy to make it go, rather than producing energy. Deuterons are a key step in burning hydrogen to helium. Without them, hydrogen would not burn, and there would be no long-lived stars and no stellar nucleosynthesis to produce the remaining elements.

Therefore, it is necessary that
\begin{equation}
M_n-M_p \leq 2.72\:\mathrm{MeV}.
\end{equation}
This reduces the congenial region on the lower left of the physical point to 
\begin{equation}
x_3 \geq -1.4.
\end{equation}

These two conditions, thus, significantly reduce the congenial corridor from 
\begin{equation}
-12.9\leq x_3 \leq 4.1
\end{equation}
to
\begin{equation}
-1.4\leq x_3 \leq -0.5.
\end{equation}
It may be noted here that the width of this region is of the same order as the uncertainty in $x_3^\oplus$ itself due to uncertainties in the light quark masses given by $x_3^\oplus =-1.17\pm 0.43$.

In fact it was a pleasant surprise to realise, rather late into our work, that Hogan \cite{Hog00} reached essentially similar conclusions which were expressed in terms of the up-down quark mass difference, $\delta m_{d-u}$ and Section IV of his review \cite{Hog00} is a recommended read for anybody interested in this issue.  The approximately 1.4 + 0.8 = 2.2 MeV window of variation that we find is in agreement with the allowed region in Fig.~1 of the same \cite{Hog00}.

\subsection*{Case III: Left half of the triangle}

If we move to the left half of the triangle, we essentially replace the down quark with a strange quark.  We know that the $s$-quark is, in some ways, like a heavy $d$-quark.  In the left half of the triangle the $s$-quark is light and the $d$-quark is heavy.   As if they simply interchange positions.  That is why Jaffe {\emph et al.} \cite{Jaf09} seem to find a symmetric congenial region in the left of the triangle.  The discussions for Cases I and II narrow it down, but do not remove it.

However, let us now turn towards the coupling between $u$-$d$ and $u$-$s$.  The $u$-$d$ coupling is much stronger, whereas the $u$-$s$ coupling is suppressed.  This is described by the well-known Cabibbo angle $\theta_C$. Where the $u$-$d$ coupling carries a factor $cos\,\theta_C$ and the $u$-$s$ coupling carries a factor of $sin\,\theta_C$, the Cabibbo angle being about 13 degrees.  

\begin{figure}
\includegraphics[width=0.8\columnwidth]{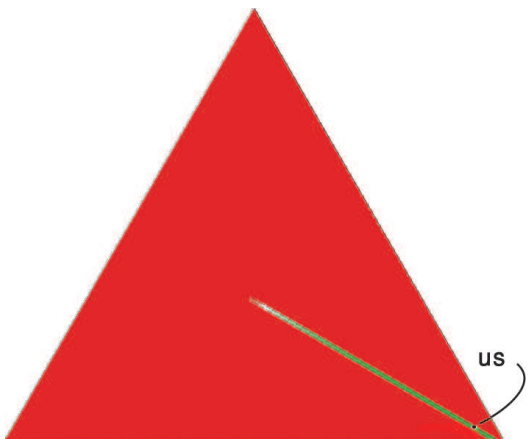}%
\caption{\label{newcong} (color online) Fig.~\ref{cong} adapted by the further restrictions imposed leaving only a very narrow congenial slit in the bottom-right region of the triangle.}
\end{figure}

This is like the present world with a much weaker weak interaction.  This the case where the weak decay rate of neutrons is not strong enough to produce the primordial neutron-proton abundance ratio of 1:6.  Without this we are left without enough protons, {\it i.e.} without enough hydrogen, which is key to both stellar burning and biological life itself.  Therefore we are left with only a narrow region on the right [Fig.~\ref{newcong}].  

The only remaining question is probably regarding the length of this narrow region extending nearly up to the centre of the triangle.  As one moves up this narrow slit towards the centre, away from the physical point, the up-down quarks become heavier keeping the down quark slightly heavier than the up. Meanwhile the strange quark becomes lighter to keep $m_T$ fixed.  The physics considered here is probably not very sensitive to the strange quark mass.  The increase in the up-down masses is offset by the compensating adjustment in $\Lambda_{\mathrm{QCD}}$ to keep the nucleon masses fixed.  Therefore, the length of this region could probably be an artifact of the simultaneous and compensating tuning of quark masses and $\Lambda_{\mathrm{QCD}}$.  

This indeed has been one of the conclusions in \cite{Jaf09} as summarized more elegantly in \cite{Jen10}; as well as \cite{Brad11} where, reviewing the alternative universe landscapes studied by \cite{Agu01,Har06,Ada08,Jaf09,Jen10} it has been observed that if one is prepared to adjust another parameter in a compensating manner, it might be possible to find other regions in the parameter space that are also congenial.  However, that does not remove the fine-tuning problem, as the alternative values are still finely tuned and this is inevitable to produce complexity as observed in our present universe.  Here most of the alternatives are removed and the narrow region remains as a result of the compensating adjustments of $\Lambda_{\mathrm{QCD}}$.

Indeed along the narrow region the sum of the two lightest quarks vary with the strange quark mass going in the opposite direction to keep $m_T$ fixed.  If the effect of this could be quantified, it would probably be possible to restrict even the length of the narrow region [See the Addendum]. 

For example, as noted by Hogan \cite{Hog00}, 
\begin{quote}
...~the sum of the (up and down) quark masses controls the pion mass, so changing them alters the range of the nuclear potential and significantly changes nuclear structure and energy levels. Even a small change radically alters the history of nuclear astrophysics, for example, by eliminating critical resonances of nucleosynthesis needed to produce abundant carbon. \footnote{An interesting additional note is that, here Hogan cites Hoyle, F., D.~N.~F.~Dunbar, W.~A.~Wenzel, and W.~Whaling, 1953, Phys. Rev. {\bf 92}, 1095.  This has been cited several times in different papers, sometimes with Phys. Rev. Lett. as the source, and a few times with the title ``A state in C12 predicted from astrophysical evidence''.  However, we failed to find any such article and would appreciate any information on this reference.  The nearest match was D.~N.~F.~Dunbar, R.~E.~Pixley, W.~A.~Wenzel, and W.~Whaling, 1953, Phys. Rev. {\bf 92}, 649, an article on the resonance in $^{12}$C often dubbed the `Hoyle resonance'.} 
\end{quote}
Here it should be added that a more up-to-date view is that the strongest effect on the scalar scattering
lengths and deuteron binding energy seem to be due to the “sigma-resonance” exchange (or correlated
two-pion scalar-isoscalar exchange) dependence on $m_\pi$ \cite{Han08,Pel10}.

As mentioned at the outset, the analysis here has been limited to the case of $m_T = m_T^\oplus$.  It has been noted by Jaffe {\it et al.} \cite{Jaf09} that the widths of the two major congenial bands on the bottom-left and bottom-right of the triangle are independent of $m_T$.  Therefore, naturally the further exclusions for $x_3 > x_3^\oplus$, $x_3 < x_3^\oplus$ reducing the width of the band should also apply to other values of $m_T$. The exclusion of the left half of the triangle should also extend to other values of $m_T$.  Therefore, in summary, it can be expected that for all values of $m_T$, after applying constraints from nucleosynthesis, there will only remain a similar very narrow congenial band at the bottom-right of the triangle. 
  
An additional comment is due here on the possibilities of universes with deuterons, sigma-hydrogen, or delta-helium playing the roles of hydrogen as listed in \cite{Jen10} as a summary of \cite{Jaf09}.  The point made here does not contradict that these could be stable lightest elements. It is only pointed out that stability alone is not enough to produce and sustain nuclear chemistry in a manner familiar to us.  Correct primordial abundances and conditions for sustained stellar burning provide constraints that are much more difficult to satisfy.  This probably calls for a closer analysis of the other half in \cite{Jen10} related to possible universes without any weak interaction of \cite{Har06} where there indeed has been a detailed discussion of these issues pertaining to nucleosynthesis.  However, that would have to be another project; whereas, this work is focused on \cite{Jaf09}.

In summary, it can be observed that, primordial nuclear abundances and processes of stellar nucleosynthesis provide much more stringent constraints on quark masses than nuclear stability.  Using these constraints it is possible to significantly reduce the congenial region in the space of light quark masses.

\section*{Addendum}
Our attention has been drawn through referee comments to studies of the bounds from nucleosynthesis \cite{Bed10,Ber13}, the latter appearing after the initial submission of this paper, on $\delta m_q /m_q$, where $m_q$ is the average of the light (up and down) quark mass and $\delta m_q$ is the change in $m_q$ keeping $m_u/m_d$ fixed.  Coincidentally, along the length of the remaining narrow congenial region $m_u/m_d$ is approximately constant.  The latest value is $\left|\delta m_q /m_q\right|<0.009$ \cite{Ber13}.  There are other values in literature, but they are generally of the same order.  Let us try to do a crude estimate of the effect of this constraint.  For $m_q\approx 3.8$ MeV, $\delta m_q\approx 0.035$ MeV.

From eqs.~8 and 9, we get
\begin{equation}
M_N=\left(M_n+M_p\right)/2=C_0+3.68\: x_T+3.53\: x_8+2.5,
\end{equation}
which given that $x_T$ is kept fixed, leads to 
\begin{equation}
\delta M_N=\delta C_0+3.53\: \delta x_8.
\end{equation}
Now, $x_8$ as defined in eq.~2 can be re-expressed in terms of $m_q$ as
\begin{equation}
x_8 =\frac{200\left(m_q-m_s\right)}{\sqrt{3}m_T^\oplus}.
\end{equation}
If $m_T$ is kept fixed then $\delta m_s=-\delta m_q$, leading to
\begin{equation}
\delta x_8 =\frac{400\left(\delta m_q\right)}{\sqrt{3}m_T^\oplus}=0.08,
\end{equation}
where we have used $\delta m_q\approx 0.035$ MeV and $m_T^\oplus\approx 100$ MeV.  The remaining congenial region is too small to show on a figure of this scale.  In fact the region $x_8=x_8^\oplus\pm 0.08$ is too small to show on a plot of this scale and is also very small compared to the uncertainty in the value of $x_8^\oplus$ itself $x_8^\oplus =-59.5\pm 1.1$ due to the uncertainties in the determination of the light quark masses.  However, one should remember our estimate is rather crude without appropriate consideration of the uncertainties in $m_q$.  Taking these into account will increase the region, but keep it within the same order as the uncertainty in $x_8^\oplus$ itself.  In short there is practically no congenial region outside $\left( x_3^\oplus ,x_8^\oplus\right)$.

\begin{acknowledgments}
The authors are grateful to Robert L.~Jaffe, Alejandro Jenkins and Itamar Kimchi for their kind replies to our queries and helpful suggestions.  MHA and ASBT would also like to acknowledge the support of the Abdus Salam International Centre for Theoretical Physics (ICTP), Trieste, Italy through a Regular Associateship and a Junior Associateship, respectively.
\end{acknowledgments}

\bibliography{CongenialityNucleosynth}
\end{document}